\documentclass[]{aastex61}

\usepackage{natbib}
\usepackage{graphicx}
\usepackage{rotating}
\usepackage{epsfig}
\usepackage{epstopdf}
\usepackage{enumitem}
\usepackage{amssymb,amsmath}
\usepackage{soul}
\usepackage{color}

\newcommand{\Rs}{$ R_{\odot}$}

\newcommand{\de}{$^{\circ}$}

\newcommand{\kms}{~km$s^{-1}$}

\newcommand{\dns}{DN $s^{-1}$}
\newcommand{\app}{$\approx$}
\makeatletter

\newcommand{\Rmnum}[1]{\expandafter\@slowromancap\romannumeral #1@}

\makeatother

\begin{document}

\shorttitle{The magnetic topology of the quiet corona}
\shortauthors{Morgan \& Korsos}
\title{Tracing the magnetic field topology of the quiet corona using propagating disturbances}
\author{Huw Morgan}
\author{Marianna Korsos} 
\affil{Department of Physics, Aberystwyth University, Ceredigion, Cymru, SY23 3BZ, UK}
\email{hmorgan@aber.ac.uk}

\begin{abstract}
The motion of faint propagating disturbances (PD) in the solar corona reveals an intricate structure which must be defined by the magnetic field. Applied to quiet Sun observations by the Atmospheric Imaging Assembly (AIA)/Solar Dynamics Observatory (SDO), a novel method reveals a cellular network, with cells of typical diameters 50\arcsec\ in the cool 304\AA\ channel, and 100\arcsec\ in the coronal 193\AA\ channel. The 193\AA\ cells can overlie several 304\AA\ cells, although both channels share common source and sink regions. The sources are points, or narrow corridors, of divergence that occupy the centres of cells. They are significantly aligned with photospheric network features and enhanced magnetic elements. This shows that the bright network is important to the production of PDs, and confirms that the network is host to the source footpoint of quiet coronal loops. The other footpoint, or the sinks of the PDs, form the boundaries of the coronal cells. These are not significantly aligned with the photospheric network - they are generally situated above the dark internetwork photosphere. They form compact points or corridors, often without an obvious signature in the underlying photosphere.   We argue that these sink points can either be concentrations of closed field footpoints associated with minor magnetic elements in the internetwork, or concentrations of upward-aligned open field. The link between the coronal velocity and magnetic fields is strengthened by a comparison with a magnetic extrapolation, which shows several general and specific similarities, thus the velocity maps offer a valuable additional constraint on models.  
\end{abstract}


\section{Introduction}
Due to observational limitations, there are large uncertainties of the coronal magnetic field topology above the quiet Sun. Reliable estimates of the magnetic field strength from spectropolarimetrical observations currently depend on exceptional observations of high signal \citep{kuridze2019}, and are particularly challenging for the quiet corona \citep[e.g.][]{khomenko2003}. Within active regions (AR), systems of large and bright loops can be observed in the extreme ultraviolet (EUV) or X-ray, providing a comparison with magnetic models, including quantitative constraints in some cases \citep{aschwanden2013}. Above the quiet Sun, the structure of the low corona as observed in EUV is bereft of distinct and bright coronal loops, leading to uncertainty. 

Insight into the quiet Sun coronal topology and connection to the lower atmospheric layers is dependent largely on model extrapolation, constrained by observations of the photospheric and chromospheric magnetic fields and their time evolution \citep{rubio2019}. Convective flows of photospheric granules are observed to form coherent supergranular cells of motion with typical diameters of 32Mm (45\arcsec) \citep[e.g.][]{tian2010,noori2019}.   A comprehensive review is given by \citet{rincon2018}  . Their boundaries are  regions of convective downflow that form lanes of magnetic concentrations that coincide with the bright chromospheric network observed in the UV continuum. \citet{schrijver2003} showed, using arguments based on a potential field magnetic model, that around half of the field may be rooted in the darker internetwork regions, with the other half arising from the network and closing into the internetwork. \citet{wiegelmann2010} found that most quiet Sun magnetic loops that reach mid-chromospheric heights or higher have one footpoint in the strong magnetic fields of the network, with the other footpoint in the weaker internetwork regions. A schematic that helps visualise these connections is figure 16 of \citet{wedemeyer2009}, which also gives a comprehensive overview of the field. We also refer the reader to the detailed review of \citet{wiegelmann2014}, and section 5 in particular.

This work uses the motion of faint propagating disturbances (PD) in time series of EUV images to gain observational insight into the distribution of the quiet Sun coronal magnetic field. The Time Normalized Optical Flow (TNOF) method to enhance and characterise the PD motions was first described by \citet{morgan2018}, with a more simple and efficient approach described by \citet{morgan2022}. The amplitudes of the PD are very faint - at most \app4 \dns, or less than 2\%\ of the background signal. They are ubiquitous and continuous, appearing quasi-periodically and showing a preferential direction of motion over local regions of the corona. Past studies of PD showing similar characteristics include \cite{stenborg2011}, who state that the disturbances are a coronal phenomenon that exist permanently everywhere, and \cite{wang2009} who found PD in a fan-like loop structure. The propagation velocities are on the order of tens of \kms, and have quasi-periodicities, or repetition times, of a few minutes. The PD are usually interpreted as slow magnetoacoustic waves, with some studies emphasising propagation along open magnetic field \citep{stenborg2011}. The events presented by \cite{morgan2018} were generally fainter compared to these other studies, and PDs were found in closed field structures in an active region and quiet Sun.   \citet{sheeley2014} used running difference images and time-distance maps to show `flows' in active regions and within cellular plumes. They describe the global character of these apparent flows, particularly within the context of widespread cellular plumes. We note that the propagating disturbances that they describe included events considerably brighter than those revealed by the TNOF method, and some of the examples had speeds considerably higher (\app90\kms, compared to below 50\kms\ for TNOF)  . 

\section{Coronal velocity field}
\label{vfield}
\begin{figure*}[]
\begin{center}
\includegraphics[width=14.0cm]{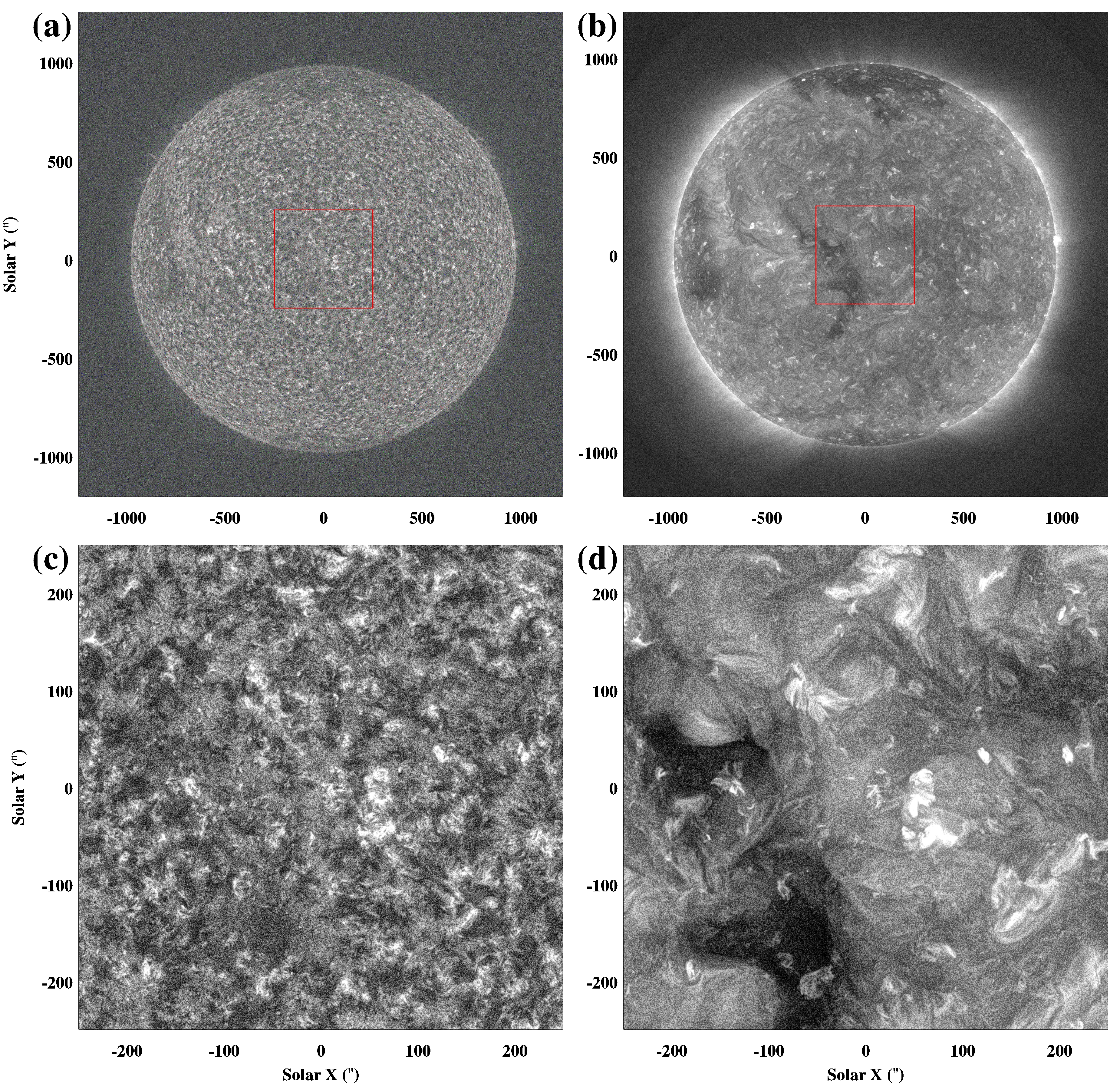}
\end{center}
\caption{Top row: The solar corona as observed by the AIA (a) 304\AA\ and (b) 193\AA\ channels at 2018 October 27 13:00UT. The red box surrounding disk center bounds the region of interest. Bottom row: Detail of the region of interest in (c) 304\AA\ and (d) 193\AA.  These images have been processed by Multiscale Gaussian Normalization \citep{morgan2014}. }
\label{context}
\end{figure*}

The TNOF method ingests a time series of solar atmospheric images and produces velocity vector maps \citep{morgan2018,morgan2022}. The method is applied here to the 304 and 193\AA\ EUV channels of the Atmospheric Imaging Assembly (AIA, \citet{lemen2011}) onboard the Solar Dynamics Observatory (SDO, \citet{pesnell2012}). The signature of faint PD are enhanced relative to the background through the use of a time-normalizing process applied to the calibrated time series signal. A Lucas-Kanade algorithm is utilised as an optical flow method to trace the motion of the PDs. In this work the period of study is between 2018 October 27 12:00UT and 14:00UT. There are no substantial eruptions or flares during this period. The choice of a two-hour period is a compromise between gaining the statistics necessary to cleanly characterise the velocity field, and minimising the effects of large-scale motions such as differential rotation, or other changes as described by \citet{morgan2022}. The region of study is within $\pm$250\arcsec\ from disk center (relative to Earth) in both $x$ and $y$, shown as the boxed red regions in figures \ref{context}a and b for 304 and 193\AA\ respectively, with more detail shown in figures \ref{context}c and d. The region is mostly quiet Sun, with   a small, decaying,   equatorial coronal hole in the south-west which appears dark in the 193\AA\ image. This dark region was a clear equatorial coronal hole in previous rotations, but by 2018 October 27 has shrunk in width. 


The movie accompanying figure \ref{vfield} shows the time series of 304 and 193\AA\ time-normalised and noise-reduced images. The movie shows complex patterns which are difficult to decipher, although they give a clear sense of consistent and repetitive motion. The Lucas-Kanade algorithm, applied to find the best least-squares fit to the $x$ and $y$ velocity components at each pixel over the 600 time steps, reveals a coherent structure that must underlie these complex patterns. These velocities are interpreted as the dominant direction and magnitude of the motions over the two hours of observation. The magnitude of the velocities range from 0 to 25\kms\ for 304\AA\, with a most probable speed of 5\kms. For the 193\AA\ channel, the speeds range from 0 to 70\kms, with a most probable speed of 15\kms. 

Figure \ref{vfield}a and b show the velocity vectors for the 304 and 193\AA\ channels respectively, superimposed on background intensity images. In the 304\AA\ velocity field, there is a general pattern of flow lines starting (red) and ending (blue) from points, or narrow corridors, of high field line density. The flow lines between these sources and sinks form cells, or sectors, of coherent flows of approximate diameters 50\arcsec. Flow lines do not necessarily form and end at the same region - flow lines from a single source can lead to two or more different sink regions. The 193\AA\ channel velocity field shown in figure \ref{vfield}b has a similar structure of coherent cells, but they are generally larger, meaning coherent flows over larger areas. Some of the 193\AA\ flow systems are considerably larger than 100\arcsec. For example, the dark coronal hole is a source of flows that reach to the neighbouring quiet Sun. 

\begin{figure*}[]
\begin{center}
\includegraphics[width=18.0cm]{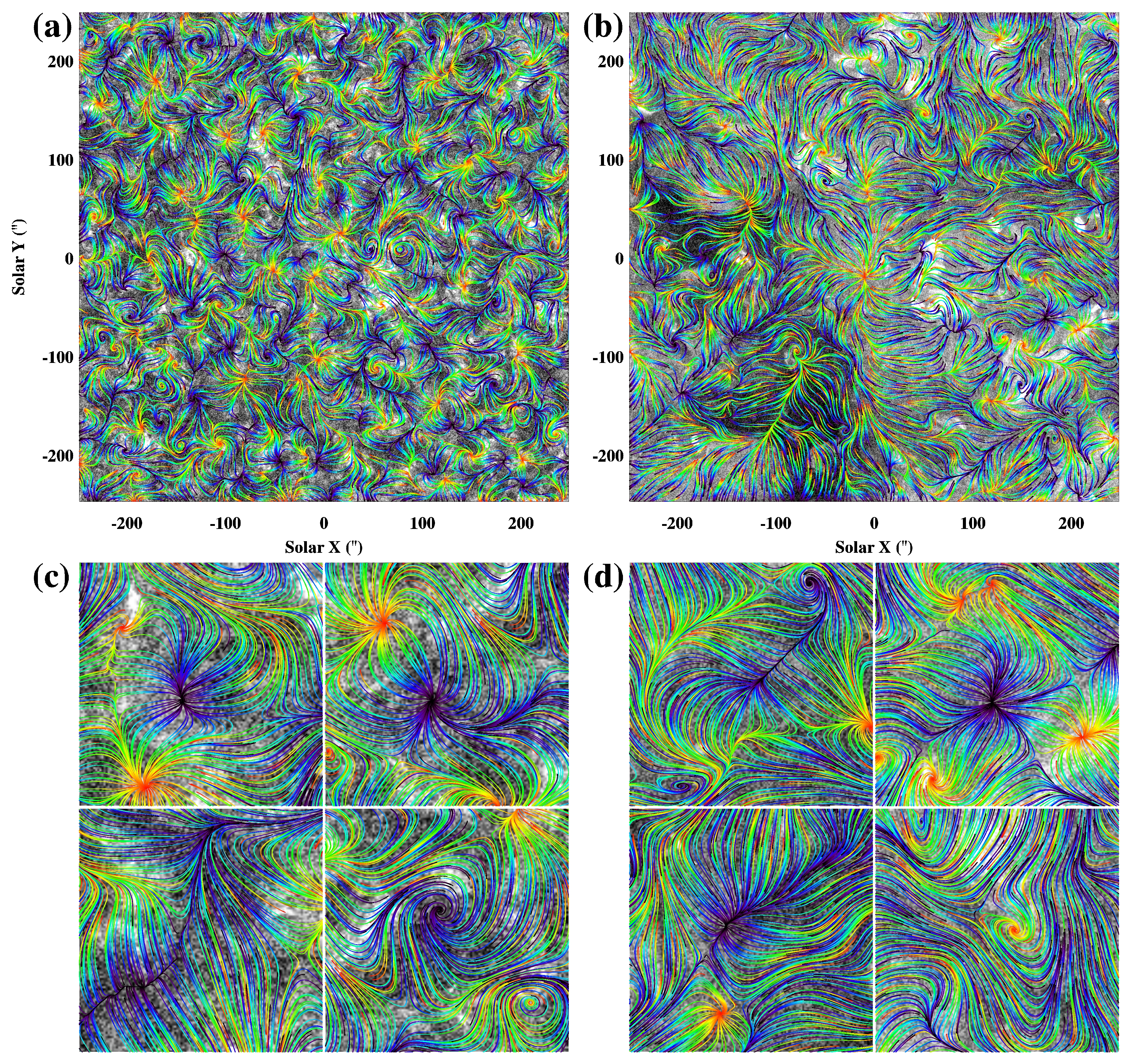}
\end{center}
\caption{Top row: Velocity vector maps, overlaid on background MGN images, for the (a) 304 and (b) 193\AA\ channels. The velocity vectors are depicted using a rainbow colour scheme, with a field line starting with colour red, and advancing through yellow, green, and ending in blue (in the order of flow direction). Bottom row: Greater detail of selected regions for the (c) 304\AA\ velocity field and (d) 193\AA\ field. For 304\AA\ in panel c, shown clockwise from top left are $70 \times 70$\arcsec\ regions centered on (10,-75), (135,-165), (-135,115), and (175, -135)\arcsec. For 193\AA\ in panel d, shown clockwise from top left are $90 \times 90$\arcsec\ regions centered on (135,105), (175,-55), (-135,-85), and (45, 145)\arcsec.}
\label{vfield}
\end{figure*}

The panels of figure \ref{vfield}c show greater detail of four selected regions of the 304\AA\ velocity field, each of size $70 \times 70$\arcsec. The first example (top left) shows a clearly-formed cell with a dense sink region in the center. The flow lines converge from all directions into this central sink. Some originate from a point source region in the lower left of the area, others arise from a dense corridor aligned north-south towards the left of the area. The top right panel shows a similar cell structure, which has a central sink region, and also a neighbouring point source region in the top left. The bottom left panel of figure \ref{vfield}c shows an example of an extended, narrow sink corridor which meanders from the bottom left to the top right corner. The bottom right panel shows a central sink region where the incoming flow lines have a consistent pattern of rotation around the central sink. In this case, the field lines describe one complete rotation from their source to their center destination. There are many examples of such rotations in the main velocity field of figure \ref{vfield}a,   and can be compared to the large distinct vortex described in section 7 and figure 20 of \citet{morgan2018}. A more detailed study may show if these rotational structures are linked to certain features on the photosphere, for example persistent rotational regions associated with supergranular vertices \citep{requerey2018}   .

The four panels of figure \ref{vfield}d show details of four selected regions from the 193\AA\ velocity field. The two top panels show two typical cell regions. The first, on the left, is a clear cell structure that has an extended sink corridor in the center, with inwardly converging flow lines. The flow lines emerge from two narrow source corridors that form the left and bottom boundary of the cell. The top right panel shows another cell example, which has a point-like sink region in the center, with three neighbouring source regions that are more localised than the extended corridors in the first example. The bottom left panel is an example of an extended corridor of convergence, and the bottom right panel is an example of a rotated source region. These rotated structures are rare in the 193\AA\ channel compared to the 304\AA.   We speculate that this is due to the corona being less closely connected to the photosphere compared to the cooler layers observed by the 304\AA\ channel, thus the smaller cell structures in the 304\AA\ channel may be closely associated with a rotating region in the photosphere, causing a degree of rotation at that height. The larger cellular structures in the 193\AA\ channel may bridge above the smaller cells of the 304\AA\ channel, and be less prone to the rotation of the photosphere at smaller spatial scales.   


In both channels, the flow lines that feed into or out of dense corridors of divergence/convergence tend to approach the corridors at a similar angle. That is, on the approach to the corridor, a flow line tends to join the corridor at an acute angle which is similar to other flow lines at that side of the corridor. Flow lines that approach from the other side of the corridor follow the same pattern, with the angle of approach reflecting the opposite side - giving a pattern akin to a fern leaf. This suggests that the flow tends to one direction along the corridor. The bottom and top left panels of figure \ref{vfield}d show examples of fern-leaf shaped divergence corridors for the 193\AA\ channel.

In the 193\AA\ channel, the flow lines within the coronal hole area (the 193\AA\ dark region in the south-east) link to neighbouring brighter regions, and in the 304\AA\ channel there is no general difference in the pattern of flow lines within this region compared to the quiet Sun. We would expect the center of equatorial coronal holes to be dominated by open magnetic field and a different flow pattern to quiet Sun regions, but the results seen here may relate to the fact that the coronal hole has decayed.   The coronal hole is discussed further in section \ref{ch}.

The top row of figure \ref{comp} shows a simplified representation of the complicated flow fields. The blue and red lines represent certain regions where the direction of the flow field changes direction abruptly. The direction of the flow field at each pixel is given by the angle $\Omega = \arctan (v_y/v_x)$. The Sobel edge enhancement operator is applied to this angular image to give a value at each pixel: very high values are found at regions where the angle in velocity direction changes abruptly - both the red and blue lines in the bottom row of figure \ref{vfield} correspond to high values. The red (blue) colours denote where these regions of abrupt angular change have positive (negative) divergence. For convenience, we denote the regions of angular discontinuities and positive or negative divergence $S^+$ and $S^-$ respectively. The blue lines overlie the points and narrow corridors where flows converge from surrounding areas (sinks), whilst the red lines overlie the points and corridors where flows diverge to surrounding areas (sources). 

\begin{figure*}[]
\begin{center}
\includegraphics[width=13.0cm]{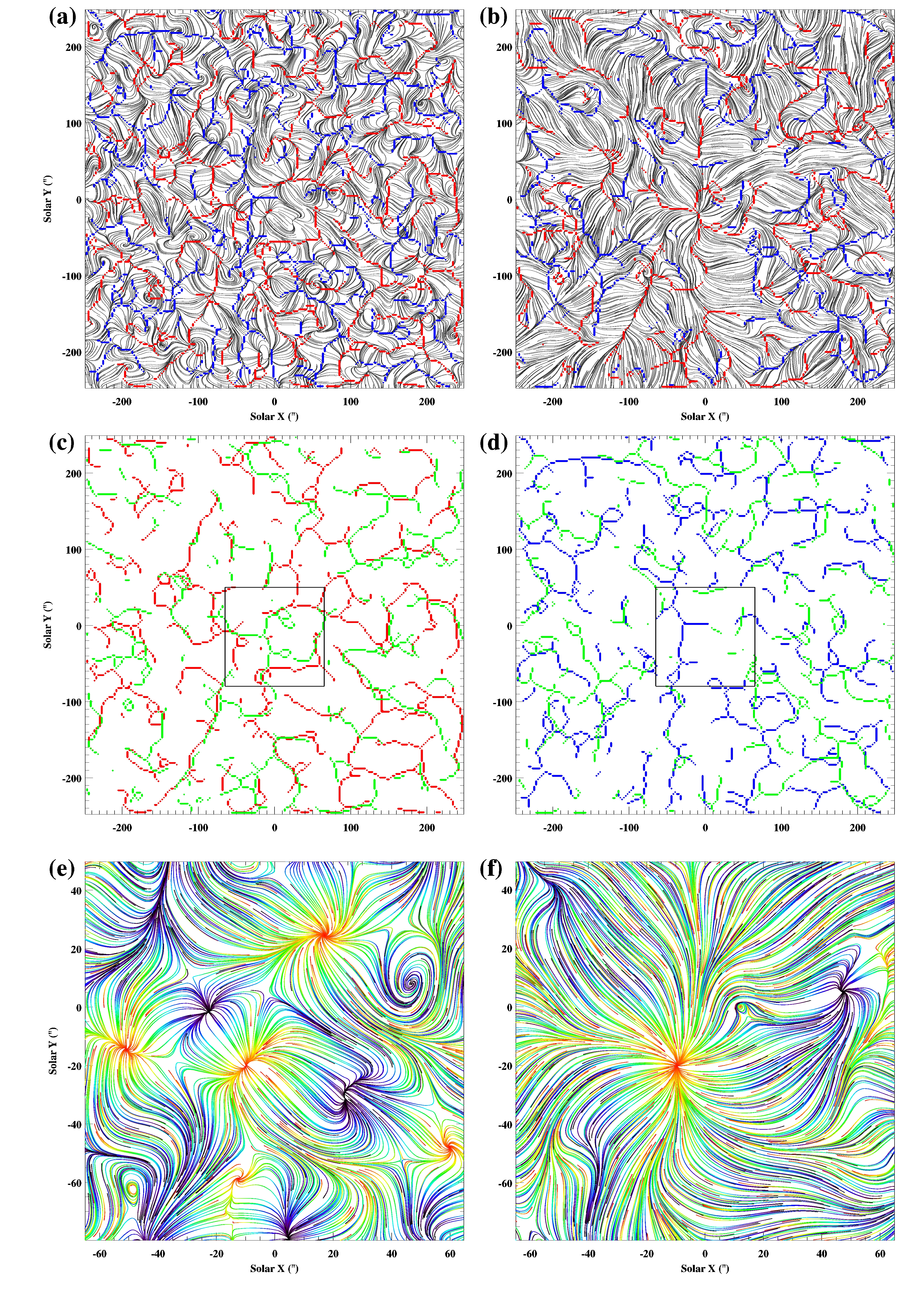}
\end{center}
\caption{Top row: Velocity vector maps shown without color or background image for the (a) 304 and (b) 193\AA\ channels. The red and blue lines show where there is a large change in direction of the local velocity vector, with red corresponding to positive divergence ($S^+$, or source of flows) and blue corresponding to negative divergence ($S^-$, or where flows converge). Middle row: (c) $S^+$ regions for the 304\AA\ (red) and 193\AA\ (green) velocity fields. These are regions of large angular change and positive divergence in the velocity vector fields, so trace out the center of flow source regions. (d) $S^-$ regions for the 304\AA\ (blue) and 193\AA\ (green) velocity fields. These trace regions of large angular change and negative divergence, or the corridors and points where flows converge. Bottom row: Further detail of the velocity vector field for the (e) 304 and (f) 193\AA\ channels, for the small region near disk center boxed in panels (c) and (d).}
\label{comp}
\end{figure*}


Figure \ref{comp}c shows the $S^+$ distribution (flow sources) for the 304\AA\ (red) and 193\AA\ (green) velocity fields. There is a correlation between the two distributions across most of the region, with a spatial alignment significance (see Appendix \ref{app}) of 2.7$\sigma$ and a percentage significance of 100\%\ (i.e. the alignment score is higher than all random cases). The 193\AA\ $S^+$ forms longer, more coherent, lines than the 304\AA, thus showing the larger-scale coherence of the 193\AA\ flow fields, and the general trend for smaller-scale structure in the cooler channel, as can be seen visually in figure \ref{vfield}a and b. In figure \ref{comp}b, which shows the $S^-$ distribution (flow sinks) for the 304\AA\ (blue) and 193\AA\ (green) velocity fields, the fields are in broad agreement in some regions, but the spatial match is overall weaker with a spatial alignment significance of 1.7$\sigma$ and percentage 95\%. Thus the cooler 304\AA\ channel and hotter 193\AA\ channels are more likely to share common source regions to their PDs, but less likely to share alignment between sink regions (or where the PD flow lines end). 

Figure \ref{comp}e and f show greater details of the velocity vector fields for 304 and 193\AA\ respectively in a small region near disk center. These figures emphasise the fact that both channels can share the same source and sink regions, but that the cooler 304\AA\ channel tends to have more source and sink regions, and has shorter flow paths with higher curvature between these regions. For example, both channels share a dominant source near the region center. They also have a sink point to the top right ($x,y=45,5$\arcsec). This sink point shows rotation only in the 304\AA\ channel. 

\section{Comparison with photospheric observations}
\label{comp}

\begin{figure*}[]
\begin{center}
\includegraphics[width=15.0cm]{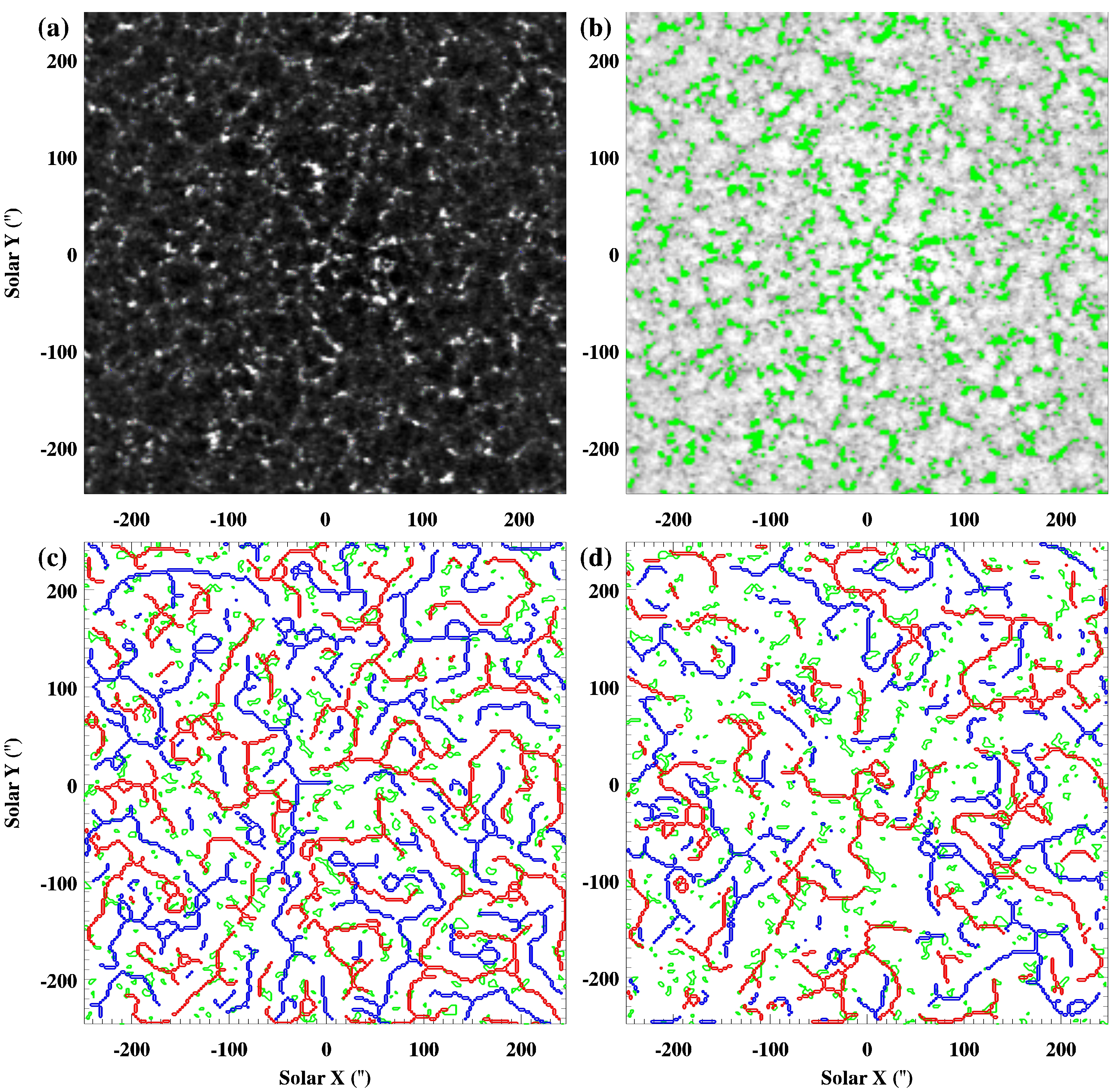}
\end{center}
\caption{(a) Observation of 2018 October 27 13:00UT of the region of study by the AIA 1700\AA\ channel, showing the photospheric UV continuum. (b) The negative of the AIA 1700\AA\ channel intensity (white = low intensity), with the brightest network features highlighted in green. (c) Comparing the UV network (green) with the AIA 304\AA\ velocity field characterisation of $S^+$ (red, positive divergence or sources) and $S^-$(blue, negative divergence or sinks). (d) As panel (c), but for the AIA 193\AA\ velocity field.}
\label{compuv}
\end{figure*}

Figure \ref{compuv}a shows an observation of the region of interest by the AIA 1700\AA\ channel, where the ultraviolet continuum is dominated by the temperature minimum/photospheric intensity. This data shows the patchwork of dark internetwork regions bounded by the bright network.   Analysis of a time series of such images taken between 2018 October 27 12:00 to 14:00UT show changes only to small-scale features, whilst the brightest large-scale network patterns remain fairly stable. The negative of the same data (black=bright) is shown as the background image of figure \ref{compuv}b, with the detected brightest network patterns highlighted in green. Figure \ref{compuv}c compares the bright photospheric network (green) with the AIA 304\AA\ velocity field characterisation of $S^+$ (red, positive divergence or sources) and $S^-$(blue, negative divergence or sinks). Visual inspection shows some areas of close agreement between red and green, thus the sources of the AIA 304\AA\ velocity field can often closely align with the photospheric bright network. The spatial alignment significance value between these is 2.8$\sigma$, with percentage score 100\%. In contrast, there is no significant agreement between the blue and green, with a percentage score of 24\%. This can be seen visually - the blue lines tend not to align over the green regions. Thus the sinks of the AIA 304\AA\ velocity field have no significant spatial correlation with the photospheric bright network. Figure \ref{compuv}d shows the same comparison for the 193\AA\ channel. As with the 304\AA\ channel, there is high significant alignment between the red and green, with a score of 2.8$\sigma$, and 100\%. There is no significance between the blue and green, with percentage 47\%. 

\begin{figure*}[]
\begin{center}
\includegraphics[width=12.0cm]{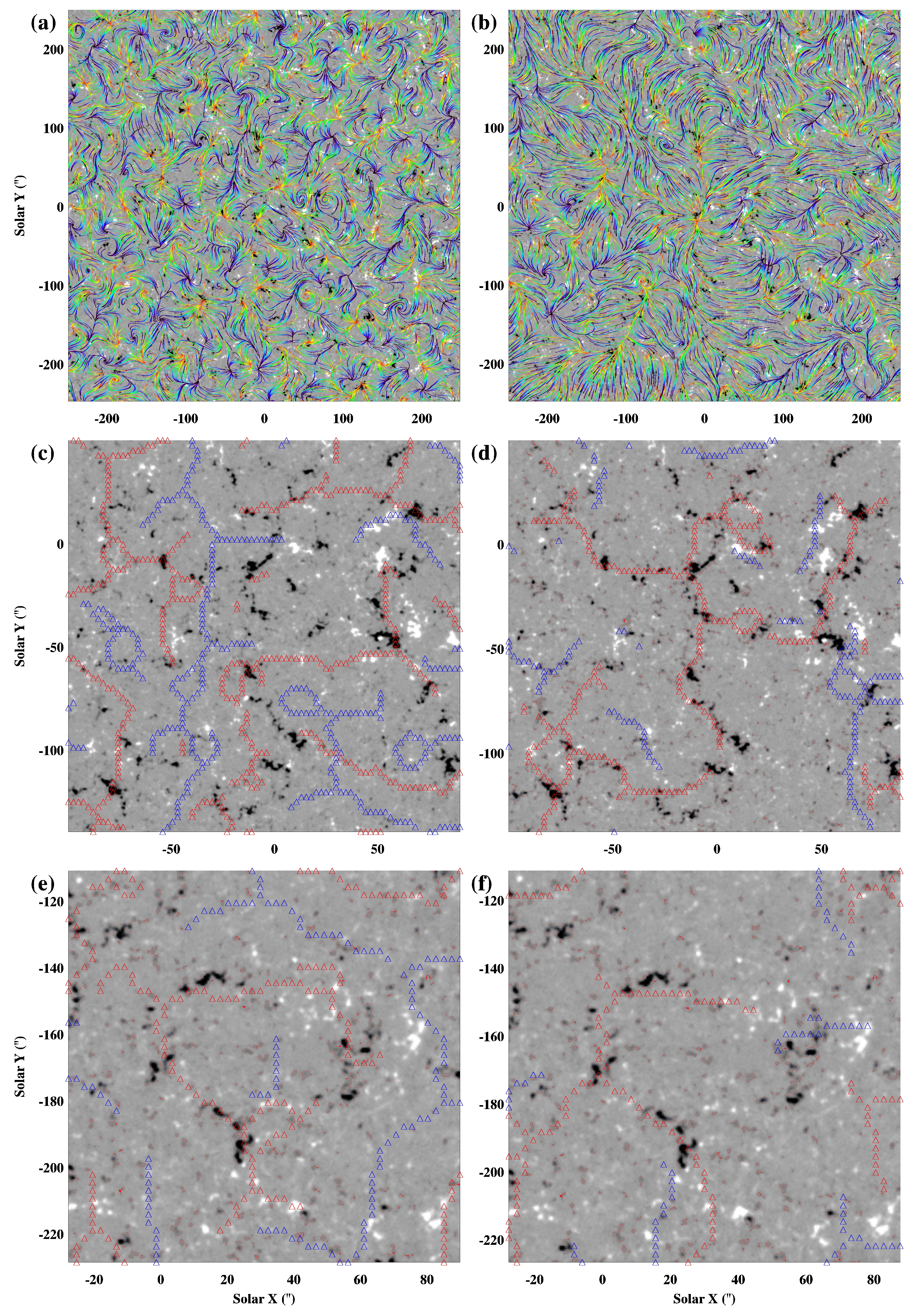}
\end{center}
\caption{A comparison of photospheric magnetograms calculated from HMI observations of 2018 October 27 13:00UT with the 304 (left column) and 193 (right column) velocity fields. The top row shows the velocity fields superimposed on the HMI magnetogram for the whole region of study. The middle row shows the source (red) and sink (blue) features of the velocity fields superimposed on the magnetogram for a small region near disk center. The bottom row shows another small region to the south of disk center.}
\label{comphmi}
\end{figure*}

Figure \ref{comphmi} shows a comparison between the 304 and 193\AA\ velocity fields and a photospheric magnetogram calculated from Helioseismic and Magnetic Imager (HMI/SDO, \citet{scherrer2012}) observations. The top two panels show the velocity vectors superimposed on the magnetogram images, whilst the bottom four panels show the source and sink features of the velocity fields for two smaller regions. Similar to the UV network presented in the previous figure, there are several regions which show close correlation between the source features (red) and features of both positive and negative enhanced photospheric magnetic field. Figure \ref{comphmi}d shows an extended 193\AA\ source feature that is closely aligned with the underlying, mostly negative, field in the HMI data. The 193\AA\ sink features are generally not closely aligned with photospheric enhancements, but can sometimes coincide. For this same region in the 304\AA\ channel, shown in panel c, there is less obvious correlation, although source features appear more likely to coincide with enhanced photospheric field compared to sinks. Figure \ref{comphmi}e shows close alignment between a loop-shaped 304\AA\ source feature and the photosphere. Most of the sink features are not aligned with enhanced field, and indeed one prominent sink feature sits within the looped source feature. A larger loop of sink points surround the inner source loop. The same region in 193\AA, shown in Figure \ref{comphmi}f, also shows close alignment between some source features and the photosphere, but the pattern is not as clear as the 304\AA\ case. 

\section{Comparison with a potential magnetic field extrapolation}
\label{comp2}
From the comparison with both the UV and magnetogram data, it is obvious that there is a link between the distribution of the coronal velocity fields and the underlying magnetic features of the photosphere. In this subsection, we extend this comparison to a potential field source surface (PFSS) extrapolation, based on the HMI magnetogram. Our code for extrapolation is based on a numerical implementation of the Green's function method \citep{sakurai1982,boocock2019}. We limit the comparison to a small region extending from -65 to 50" in both $x$ and $y$, encompassing most of the region shown in the middle row of Figure \ref{comphmi}. For efficiency, we resize the magnetogram to a quarter of the number of pixels, so that each new pixel is the local average of four pixels of the original. To avoid edge effects, the extrapolation area extends by 10" beyond the displayed area at all margins, and the photospheric input to the extrapolation extends well beyond this. The extrapolation volume extends to 0.2\Rs\ above the photosphere. We use all the magnetogram pixels to calculate the potential field, not just enhanced sources, and we do not fit the photospheric field to a smooth function. Thus the potential at a given coronal point in the extrapolation volume is calculated from all photospheric points within the visible horizon.

\begin{figure*}[]
\begin{center}
\includegraphics[width=18.0cm]{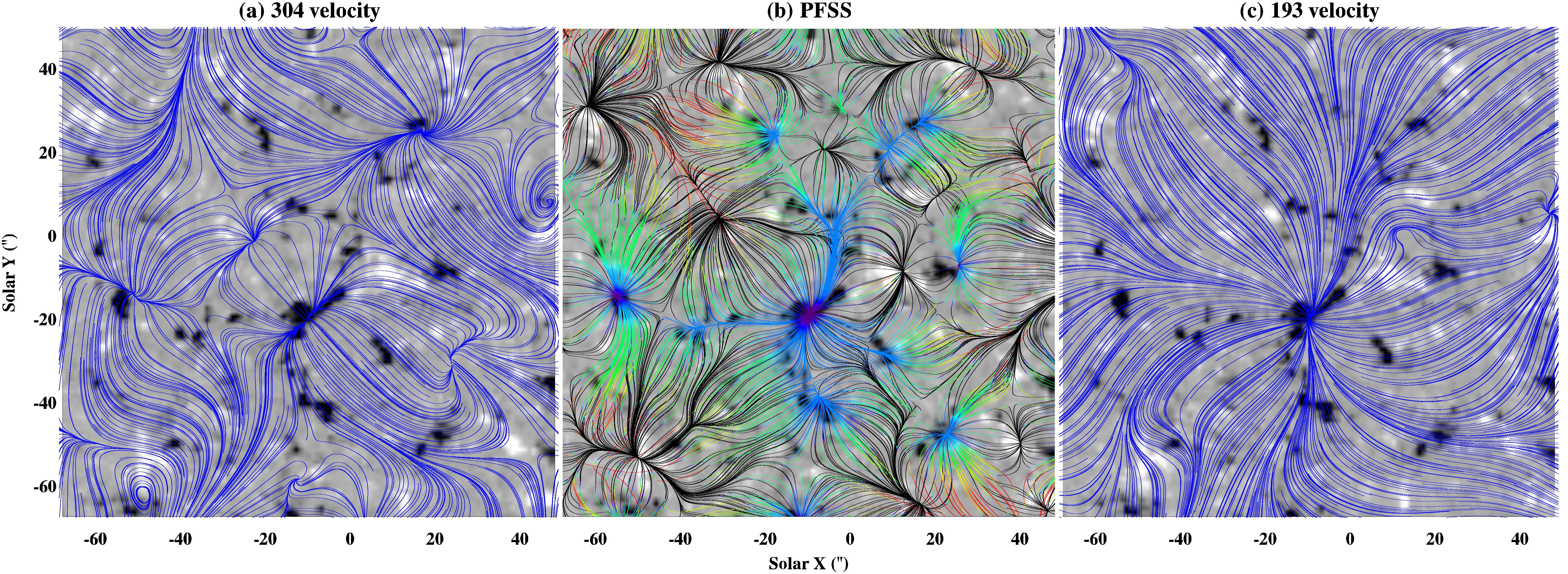}
\end{center}
\caption{Comparison of the velocity fields of the (a) 304 and (c) 193 channels, with (b) field lines traced from a potential field extrapolation, for a small region near disk center. The extrapolated field lines are coloured according to height, with black being very low-lying lines, followed by blue, green, yellow and red (near the maximum extrapolated height of 0.2\Rs). The background image is the HMI magnetogram observed at 2018 October 27 13:00UT.}
\label{comppfss}
\end{figure*}

Figure \ref{comppfss}b shows the result of the PFSS extrapolation. The field lines are traced in the 3D volume, and their coordinates transformed to the observer's field of view.  There are many similarities between the PFSS and velocity fields, as well as some major differences. On a general level, the PFSS field is cellular, with dense concentrations of footpoints aligned with stronger photospheric magnetic elements. The field lines at greater heights (coloured green/yellow/red) can bridge over closed underlying systems at low heights (coloured black). Furthermore, whilst many field lines have one footpoint in a dense source, and the other in a low-density internetwork region, lines can also join in dense, narrow corridors, similar to the fern-leaf patterns seen in the velocity fields. We can thus find most of the general features of the velocity fields in the PFSS extrapolations.

Beyond these general similarities, there are also specific features that are shared between the PFSS and the 304\AA\ velocity field, including:
\begin{itemize}
\item A very similar distribution of field lines within, and surrounding, the central prominent feature at $x,y=[-10,-20]$". This is also similar in the 193 velocity field.
\item The cell systems centered at $[-35,0]$", and $[-55,-15]$", and their connection to the central prominent feature and each other.
\item A cross-shaped meeting of field lines at the boundary of distinct cell systems at $[-10,-55]$". Two other similar systems are seen at $[0,45]$" and $[35,25]$". The corresponding features in the PFSS extrapolation are at slightly different locations.
\end{itemize}
We note that the 304\AA\ velocity field contains several rotating features: two of them are seen here centered at $[-50,-60]$" and $[45,5]$". There is no corresponding spiral pattern in the PFSS extrapolation. That there is no spiral structure in the magnetic field of the PFSS model is expected, since the model has no time-dependence, and is force-free.

\section{The coronal hole}
\label{ch}

\begin{figure*}[]
\begin{center}
\includegraphics[width=12.0cm]{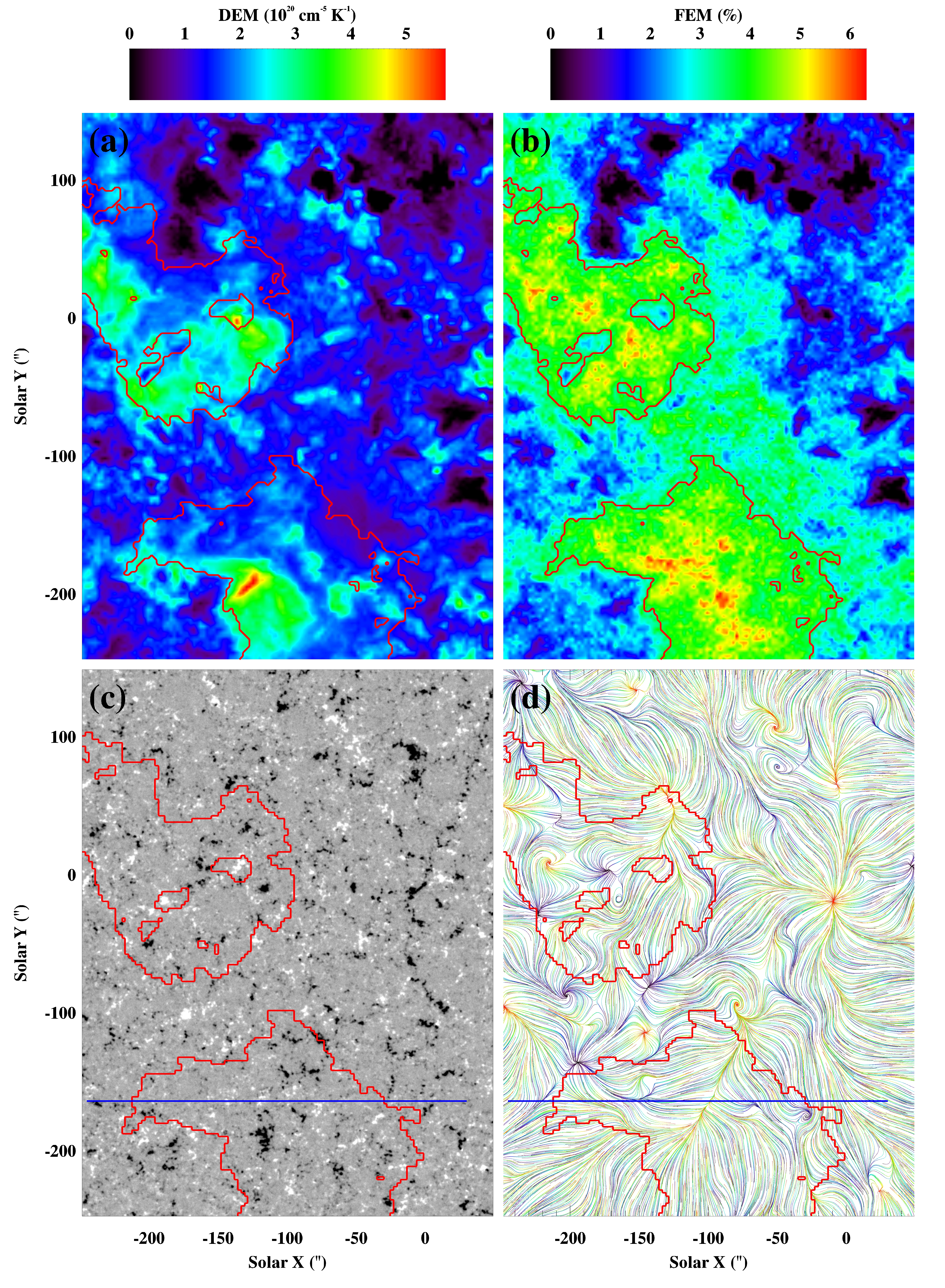}
\end{center}
\caption{A closer look at the small coronal hole. (a) A map of the differential emission measure (DEM) at a temperature very near 0.6MK. (b) The fractional emission measure (FEM, see text) at 0.6MK, or the percentage of emission at this temperature compared to the total emission over all temperatures. Values of FEM above 3.5\%\ are used to estimate the coronal hole boundary, shown in all panels as the red contour. (c) The HMI magnetogram for this region. The horizontal blue line shows a slice of the magnetogram used for the 2-dimensional potential field extrapolation shown in figure \ref{pfss2d_ch}. (d) The 193 velocity field for this region.}
\label{dem}
\end{figure*}

This section makes a more detailed analysis of the narrow coronal hole seen in the south-east of the region of study. In particular, we expect coronal holes to be dominated by open field, and further expect that field to be expanding super-radially from the central regions of the coronal hole, which seems to be in contradiction to the velocity field patterns seen in figure \ref{vfield}a and b, and shown in greater detail in figure \ref{dem}d for the 193\AA\ channel.

\begin{figure}[]
\begin{center}
\includegraphics[width=8.0cm]{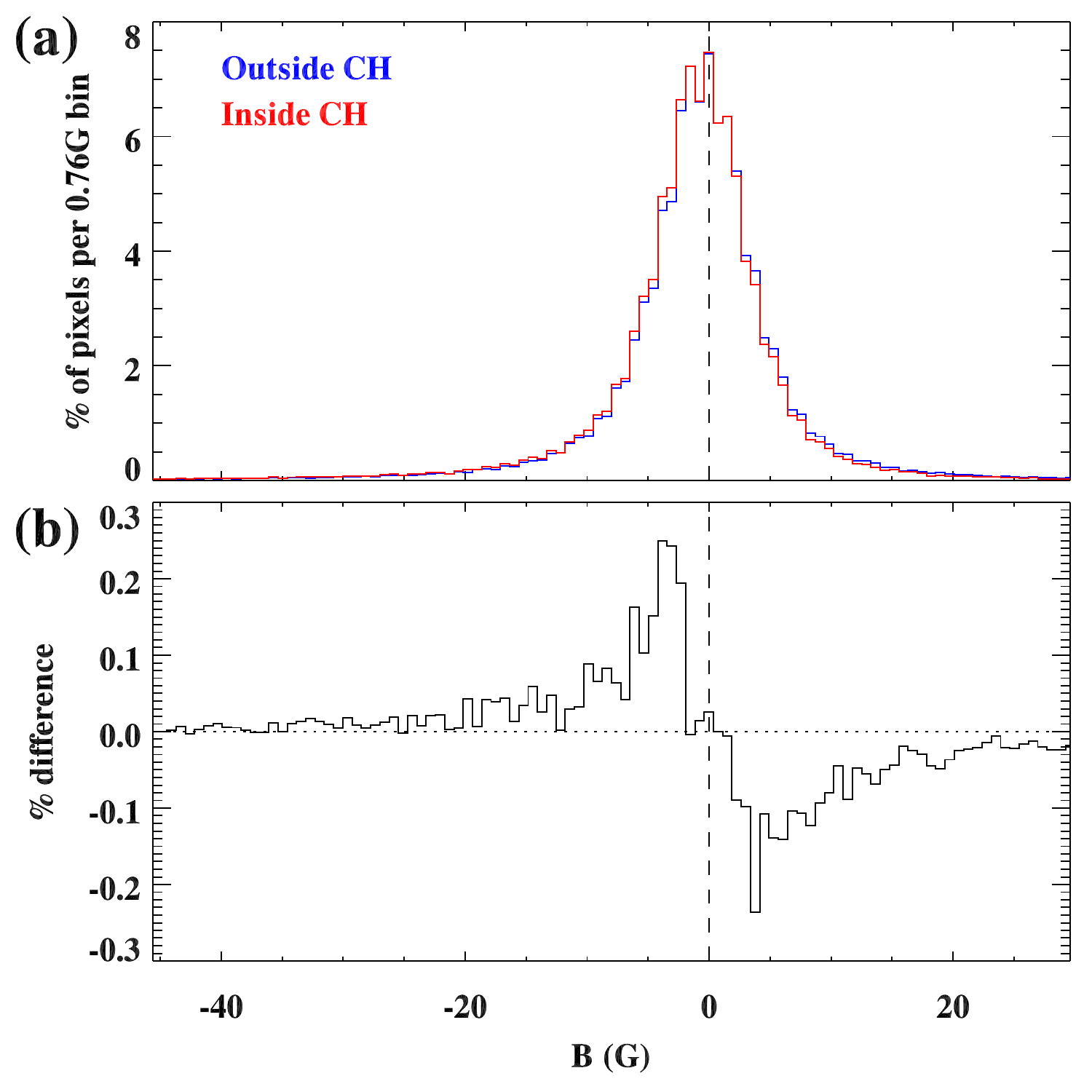}
\end{center}
\caption{(a) Histograms of the photospheric magnetogram values outside (blue) and inside (red) the coronal hole, shown as a percentage per 0.76G bin of the total number of pixels. (b) The difference between the two histograms (inside minus outside).}
\label{maghist}
\end{figure}

We first define a boundary to the coronal hole using a differential emission measure (DEM) analysis. The method we use is Solar Iterative Temperature Emission Solver (SITES, \citet{morgan2019,pickering2019}), and the input data is AIA/SDO data from the seven EUV channels. The resulting DEM, at a cool temperature of 0.6MK, is shown in figure \ref{dem}a. Figure \ref{dem}b shows the fractional emission measure (FEM), which is the emission at that temperature divided by the total emission summed over all temperatures. The map shows a clear region of enhanced FEM within, and surrounding, the coronal hole. The hole boundary, shown as the red contour, is defined by any regions that have an FEM over 3.5\%, and with the number of connected pixels larger than a few hundred. Figure \ref{dem}c plots the coronal hole boundary over the HMI magnetogram. Visually, there is no discernible difference between the photosphere within the hole, and outside in the surrounding coronal hole. Figure \ref{maghist}a shows that the histograms of photospheric magnetic field for inside and outside the coronal hole are very similar - indeed, their mean and standard deviations are close to identical. It is only when one histogram is subtracted from the other that we see a small bias towards negative polarities within the coronal hole, as shown in figure \ref{maghist}b. 

\begin{figure*}[]
\begin{center}
\includegraphics[width=14.0cm]{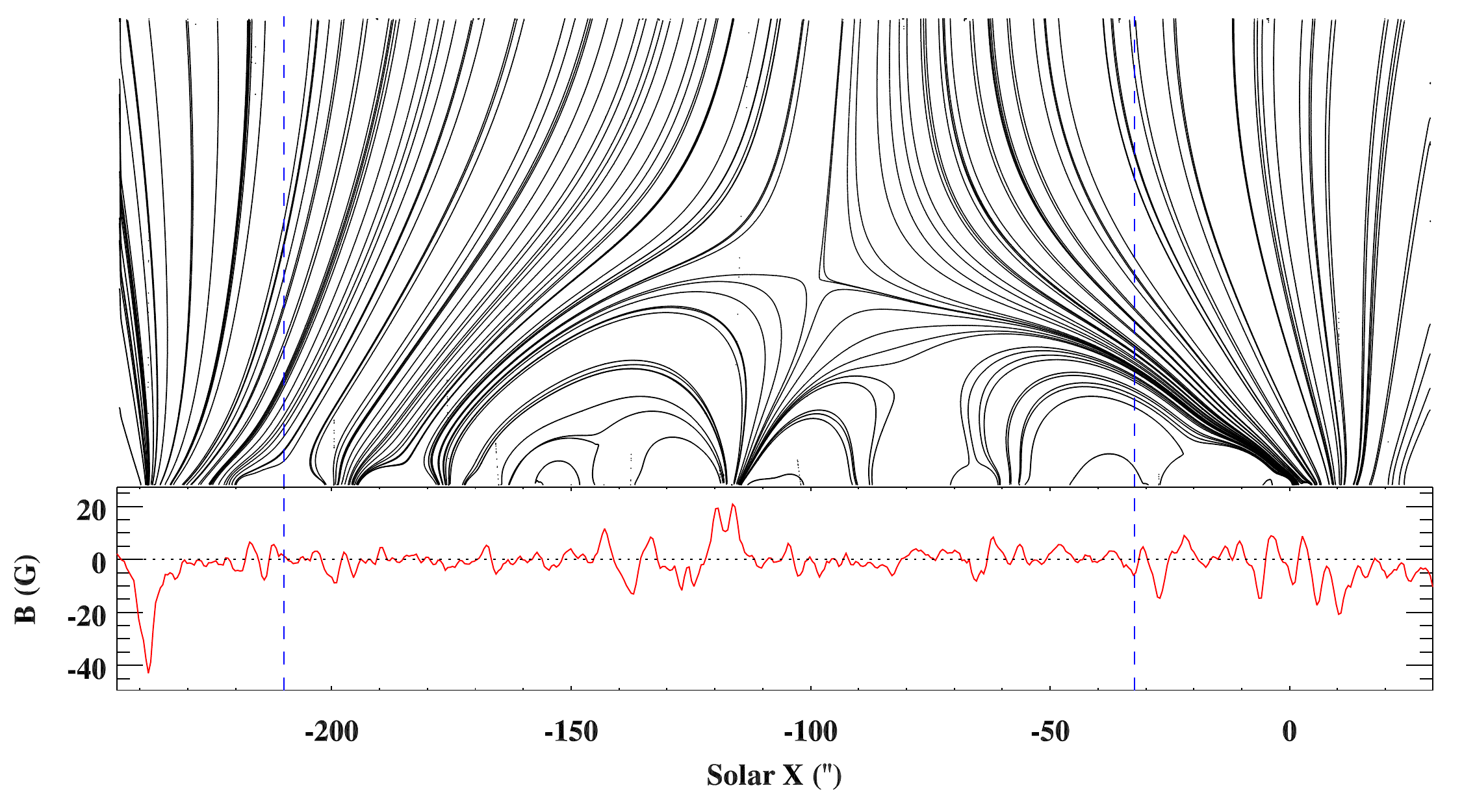}
\end{center}
\caption{The bottom panel shows the values extracted from the HMI magnetogram along the horizontal slice shown as a blue line in figure \ref{dem}c. The top panel shows the potential field extrapolation (two-dimensional: $x$, and height), based on these values. The two vertical blue dashed lines show the boundaries of the coronal hole - areas between these lines are within the coronal hole.}
\label{pfssch}
\end{figure*}

We extract the photospheric field values for a line cutting east-west across the coronal hole, shown as the horizontal blue line in figures \ref{dem}c and d. This field is shown in the lower panel of figure \ref{pfssch}. We calculate a PFSS extrapolation for this field, in two dimensions ($x$, and height), which is displayed in the top panel of figure \ref{pfssch}. Whilst a 2-dimensional extrapolation cannot be used to study the coronal hole magnetic field properly, this example helps to illustrate an argument that the velocity field may be giving a correct approximation to the magnetic field. This coronal hole does not have a large dominant polarity, and contains both positive and negative enhancements. In particular, the positive enhancement seen here in the centre of the hole, gives rise to a high density of field lines, from two system of closed structures to both east and west. Open field bounding these closed systems bridge across, and angle sharply upwards approximately above the positive enhancement. We argue that the velocity field within the southern region of the coronal hole, as seen in the 193\AA\ channel (figure \ref{dem}d), may arise from such a configuration. We note that this coronal hole is decaying, and our results cannot be generalised for all small equatorial coronal holes.

\section{Discussion \& Summary}
\label{discussion}

The flow maps are tracing the dominant direction of motions of faint, ubiquitous PDs in the solar atmosphere. The PDs must follow the direction of the underlying magnetic field \citep[e.g.][]{stenborg2011,wang2009,sheeley2014}. This is the simplest interpretation of the patterns we see in the velocity flow maps, and of the differences and similarities between the results for the 304 and 193\AA\ channels. The PD velocity fields are tracing the plane-of-sky projection of the magnetic field at the atmospheric height corresponding to the  dominant   temperature of formation of that channel,  accepting that (i) both channels contain contributions from multiple emission lines thus have a range of formation temperatures, and (ii) the height varies in different atmospheric structures . We of course lack the three-dimensional (or line of sight) information; we have only the projected (or plane-of-sky) velocities. Nevertheless, the flow lines give a unique proxy to the quiet Sun coronal magnetic field structure based directly on EUV coronal observations - information that is currently impossible to gain without magnetic modelling. The pattern seen in the quiet Sun region of figure \ref{vfield}, of cells, or sectors, of flow line systems arising from and ending in points and narrow corridors, leads to a hypothesis that the magnetic field topology has a similar sectorial structure. 

In the lower atmosphere, corresponding to the 304\AA\ channel, we expect the closed magnetic structures that form the quiet Sun corona to have generally smaller structural scales than the overlying coronal field traced by the 193\AA\ channel, and this is what is seen in the PD velocity fields. In the cooler 304\AA\ channel, the cell diameters are typically \app$50$\arcsec, whilst the hotter 193\AA\ channel tends to have larger cell sizes, with some larger cells of diameter 80 to larger than 100\arcsec. The coronal field traced by the 193\AA\ channel can share common sources and sinks with the 304\AA\ channel, but can often bridge over one or more cells in the cooler 304\AA\ channel. 

The cells tend to have central point-like, or corridor-like, source regions, with neighbouring systems of sink points or corridors forming their boundaries. Some cells show clear rotational motion. These are common in the 304\AA\ channel, and rare in the 193\AA\ channel. The apparent rotational flows suggests that the magnetic field may be twisted due to the coherent large-scale rotation of areas of the underlying photosphere. The projected fern-leaf shape of narrow sink/source corridors suggests that there is a dominant direction of flow along the narrow corridors, and a dominant orientation of the magnetic field along the narrow corridors.

The comparison with the underlying photospheric UV intensity, and with photospheric magnetograms, shows that the sources of the field lines tend to overlie the bright network, whilst their sinks tend to lie in the darker internetwork regions. This result is important. It reinforces the hypothesis that the velocity fields are a tracer of the magnetic fields, and is consistent with the accepted model of the bulk of the quiet Sun magnetic field concentrated in network regions. Furthermore, if the PDs are slow magnetoacoustic waves, then these brighter UV network regions, and enhanced magnetic elements, are conducive to the production of waves. If the internetwork regions were equally productive, then we would not see a clear preferential direction of the PD flows from the network and towards internetwork regions. Certainly there is much further work that can be done to investigate this relationship, including a more detailed comparison with the photospheric and lower-chromospheric magnetic field distribution, and a comparison with the results of motion tracking in the photosphere \citep[e.g.][]{chian2019}.  


The most surprising result is the distribution of the flow sinks (convergence regions). That the flow sources tend to emerge from dense, narrow points or corridors, and that these tend to align with bright network regions, is consistent with general models of the quiet Sun magnetic field. The source field arising from the network boundaries is expected to expand with altitude, filling the transition region and low corona, with the closed field forming the basic building blocks of the quiet Sun corona. However, our results show that the opposite footpoints (or sink footpoints) of these closed loop systems are also distributed in dense narrow points or corridors, similar to their source footpoints. That is, the field arising from the network does not spread to sink footpoints distributed at low density through broad regions of the larger internetwork regions. Whilst the field lines linking both footpoints expand to fill intervening areas, the footpoints diverge from, and converge to, distinct, small areas forming points or corridors. In contrast to the source regions, the sink regions are not aligned with the photospheric network boundaries, and there is not always a distinct signature of the sink regions in the UV photospheric intensity or in photospheric magnetograms - they are situated within the weak field of the dark internetwork regions.

\begin{figure*}[]
\begin{center}
\includegraphics[width=17.0cm]{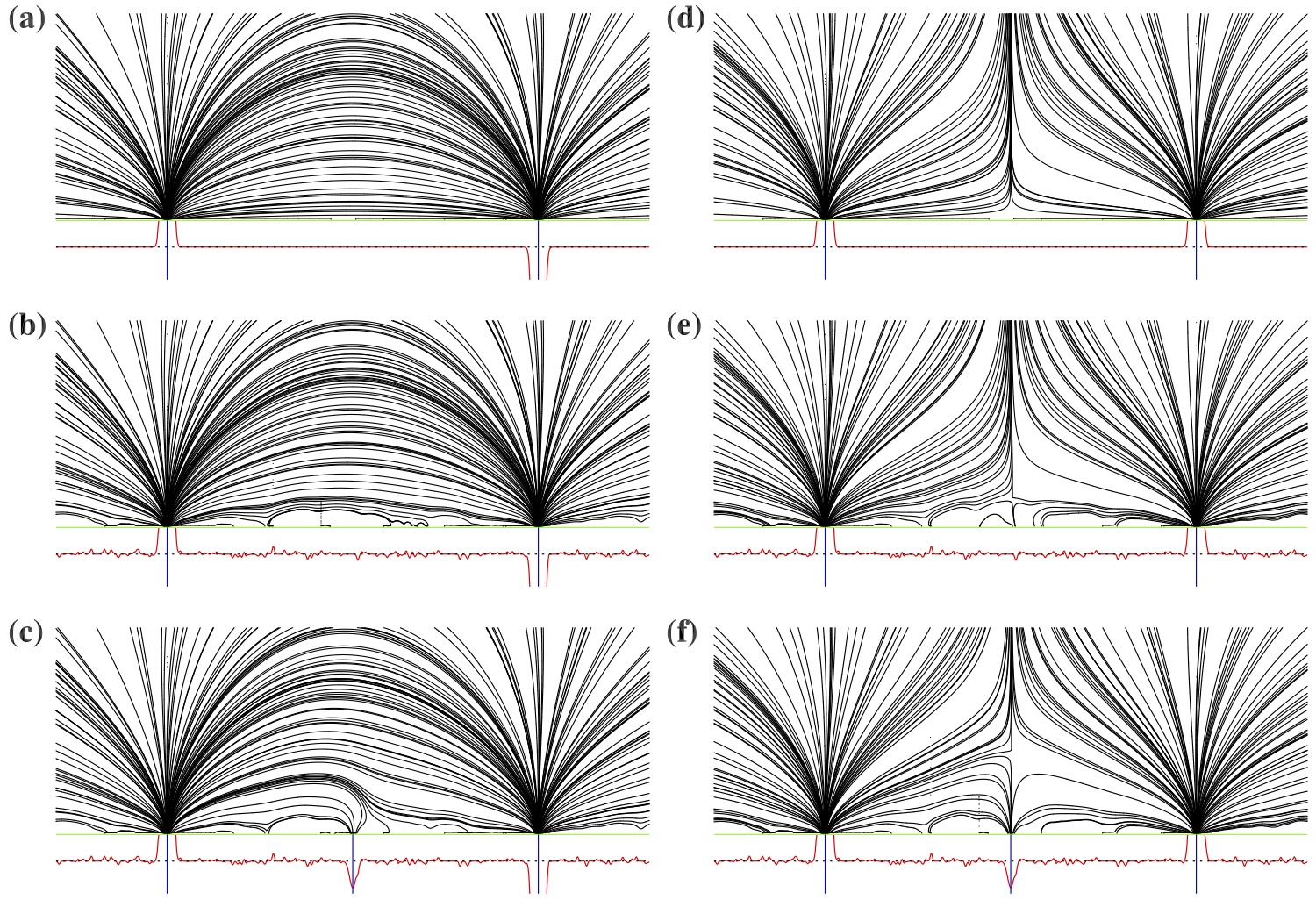}
\end{center}
\caption{Experiments with a simple two-dimensional potential magnetic field. The left column shows cases with two strong magnetic sources, one positive on the left, and one negative on the right. The right column shows cases with two strong positive sources. In each panel, the green line shows the photosphere. Below this green line, the red line shows the strength of the source (photospheric) field, with the location of the two strong sources shown by vertical blue lines. The range of the source field is limited to small values around zero in order to show small variations, so that the left and right source peaks are heavily curtailed. The top row show the case where there are two strong sources, and zero field elsewhere. The middle row adds small random photospheric variations, with standard deviation amplitudes of 1\%\ of the amplitude of the two strong sources. The bottom row has the two strong sources, random variations, and a small central negative source with amplitude 6\%\ of the two strong sources. The position of this small negative source is indicated by the vertical blue line, equally spaced between the two strong sources.}
\label{pfss2d}
\end{figure*}

Support for the presence of concentrated sink points/corridors in the quiet Sun internetwork regions can be given by simple experiments with a two-dimensional PFSS model. First, we place two strong magnetic elements, of opposite sign, in otherwise zero field, resulting in the expected configuration of coronal field as shown in figure \ref{pfss2d}a. We add to this a large number of very weak small sources with random locations and random amplitudes of standard deviation 1\%\ of the large sources. Despite the weakness of this photospheric random source, the field at low heights is perturbed, as shown in figure \ref{pfss2d}b. We now add a weak negative element of amplitude 6\%\ of the strong elements, equidistant between them. The field is now quite strongly perturbed at low to medium heights, and several field lines link the strong positive element to the weak negative element, as shown in figure \ref{pfss2d}c - leading to a concentration of field lines at the position of the small negative enhancement. The experiment is repeated with two strong positive elements in the right column of figure \ref{pfss2d}. In the case of zero field, or weak random field, outside of the strong elements (figure \ref{pfss2d}d and e), the coronal field has a concentration of upwardly-pointing open field at the centre of the system. When a weak negative element is placed in the centre (figure \ref{pfss2d}f), we have a strong concentration of both downwardly- pointing closed field as well as the open field. We argue that such configurations may be common in the quiet Sun, and can explain the dense sink regions for our velocity fields. In the case of figure \ref{pfss2d}d, the central region between the two strong positive sources has a dense concentration of upwardly-aligned open field, and this would appear as a dense sink region in a coronal PD velocity field. Relatively weak enhancements in the internetwork region can also lead to concentrations of closed field lines in the internetwork, which would also appear as sink regions in the velocity fields (figures \ref{pfss2d}c and f). 

We note that a recent detailed observational study of the magnetic configuration of a supergranule by \citet{robustini2019} suggested that the chromospheric field lines tended to point inwards from the supergranule boundary, and may form a canopy above the unipolar interior region. Furthermore, they found a persistent internetwork small brightening coinciding with a distinct blue-shifted (downflow) feature in the line-of-sight velocity of the Ca II K line. \citet{chian2019} showed repelling Lagrangian coherent structures concentrated at the center of a supergranule.  Thus central regions in the internetwork are special - they are central source regions of local supergranular-scale photospheric motions, and their properties may lead to a higher concentration of overlying atmospheric magnetic field lines. 

Our next efforts are aimed to better understand the connection between the velocity fields and the atmospheric magnetic field. We will combine our PD tracking method with spectroscopic observations of EUV lines, and will focus on whether the source and sink regions coincide with regions of Doppler shifts or other distinguishing properties. We plan to create disk-center velocity fields for AIA/SDO during periods when other spacecraft hosting EUV imagers (e.g. Solar Orbiter) are at 90\de\ to Earth, enabling a comparison of the equatorial off-limb structure in EUV with the velocity field distributions from the SDO viewpoint. We are also interested in comparing our coronal flow fields with photospheric and chromospheric flow fields.   In the longer term, an exciting possibility is to use our results as a proxy for the projected plane-of-sky distribution of coronal magnetic fields, and use this as an empirical constraint on magnetic extrapolation models. This can be particularly useful for the quiet Sun and coronal holes, where the EUV emission does not trace systems of clear loops as is seen in active regions.



 \begin{acknowledgements}
We acknowledge (1) STFC grants ST/S000518/1 and ST/V00235X/1, (2) Leverhulme grant RPG-2019-361, (3) the excellent facilities and support of SuperComputing Wales. The SDO data used in this paper is courtesy of NASA/SDO and the AIA, EVE, and HMI science teams.
 \end{acknowledgements}

\appendix
\section{Quantifying the agreement between two binary spatial distributions}
\label{app}
Two-dimensional masks $M_0$ and $M_1$, both with $n$ pixels and the same shape, contain values of only zeros and ones, with the ones (zeros) corresponding to some definition of true (false). For example, the blue and green regions of figure \ref{comp}a correspond to true values for $M_0$ and $M_1$ respectively. $M_0$ is convolved with a narrow two-dimensional Gaussian kernel $k$ giving smoothed mask $S_0$:
\begin{equation} 
S_0 = M_0 \otimes k,
\end{equation}
\noindent and the same for $S_1$. The purpose of the convolution is to spread narrow structures in the mask, enabling some overlap between closely-aligned structures that may not exactly overlap in both masks. In this work, we use a sigma-width of 1 pixel for $k$. $P_{01}$ is the product of $S_0$ and $S_1$: $P_{01}=S_0 S_1$, and a mask $N_{01}$ is defined as $N_{01}=P_{01}\geq t$, where $t$ is a small threshold. In this work, $t=0.05$. Finally, a score $c_{01}$ for the agreement between the two masks $M_0$ and $M_1$ is given by 
\begin{equation}
c_{01} = \frac{2n \sum N_{01}}{\sum N_{00} + \sum N_{11}}, 
\end{equation}
\noindent where the summation is over all pixels. The denominator is a normalising factor, where $N_{00}=P_{00}\geq t$, and $P_{00}=S_0 S_0$ (and similar for $N_{11}$). The demoniator ensures that $c_{01}=1$ for the case of $M_0=M_1$, and $0 \leq c < 1$ for the general case.

A random test is used to quantify the significance of the score $c_{01}$. $M_0$ is segmented into a set of coherent regions (groups of connected pixels). Each group's centroid is translated to a random location, with the connected pixels rotated by a random angle around this new centroid. If coordinates of the translated/rotated region lie outside of the bounds of $M_0$, those coordinates are wrapped to the opposite boundary. Thus a new random mask is created which has similar characteristics to the unmodified mask, composed of the moved and rotated connected regions of $M_0$. A score $c$ is calculated between this random mask and $M_1$. This process is repeated many times (1000 for this work), with the score $c$ recorded for each case. The resulting normal distribution of $c$ allows a test of the significance of $c_{01}$ through evaluating the mean $m$ and standard deviation $\sigma$ of $c$, and calculating $c_{01}$'s distance from $m$ in units of $\sigma$. More simply, we can assign a percentage significance by counting the number of cases of the random distribution with a lower score than the actual distribution. For example, if 900 of the 1000 random cases have a lower score than the actual, then the spatial agreement percentage is 90\%.


\end{document}